\documentclass[aps,prb,twocolumn,showpacs,amsmath,amssymb,floatfix]{revtex4}
\usepackage{epsfig}
\usepackage{bm}

\begin{document}
\title{Electron-electron scattering effects on the Full Counting 
Statistics of Mesoscopic Conductors}
\author{S. Pilgram}
\affiliation{D\'epartement de Physique Th\'eorique, Universit\'e de Gen\`eve,
        CH-1211 Gen\`eve 4, Switzerland}
\date{\today}

\begin{abstract}
In the hot electron regime, electron-electron scattering strongly modifies not
only the shot noise but also the full counting statistics.  We employ a method
based on a stochastic path integral to calculate the counting statistics of
two systems in which noise in the hot electron regime has been experimentally
measured. We give an analytical expression for the counting statistics of a
chaotic cavity and find that heating due to electron-electron scattering
renders the distribution of transmitted charge symmetric in the shot noise
limit. We also discuss the frequency dispersion of the third order correlation
function and present numerical calculations for the statistics of diffusive
wires in the hot electron regime.
\end{abstract}

\maketitle

\section{Introduction}

During the last ten years, nonequilibrium noise measurements have become a
standard tool in mesoscopic physics, because they reveal additional
information about mesoscopic conductors beyond linear response\cite{Blanter1}.
Recently, even further experimental progress has been made: Reulet, Senzier
and Prober measured for the first time successfully the third moment of
current statistics of a tunnel junction\cite{Reulet1}.  They discovered a
surprising temperature dependence of the third moment which was explained by
the backaction of the resistive measurement device on the
junction\cite{Beenakker1}.

Nonequilibrium noise and third moment of current fluctuations are part of a
more general concept, the full counting statistics (FCS) which is defined as
the probability distribution of charge that passed an electric conductor
during a measurement.  The introduction of FCS into mesoscopic physics by
Levitov and Lesovik\cite{Levitov1} one decade ago has inspired a lot of
theoretical work.  Whereas early work concentrated mainly on noninteracting
systems, recent publications developed schemes which include effects due to
Coulomb interactions.  Most of these works consider correlated electron
systems in which intrinsic\cite{Andreev1,Bagrets1,Bagrets2} or
environmental\cite{Kindermann1,Kindermann2} Coulomb blockade plays a central
role.  However, Coulomb interactions are important as well for the FCS of
semiclassical systems: At low temperatures in the hot electron regime,
electron-electron scattering leads to a local thermalization of the electron
gas. Since inelastic electron-phonon scattering is strongly suppressed, the
local electron temperature may be different from the lattice temperature and
is allowed to fluctuate. These temperature fluctuations modify the intensity
of current noise and therefore influence the FCS.

Nonequilibrium noise in the hot electron regime has first been measured in
diffusive wires\cite{Steinbach1}. Since heating effects are difficult to
avoid, it only became possible later to extend these measurements to the cold
electron regime in which electron-electron scattering is absent\cite{Henny1}.
Similar noise experiments in the hot electron regime have been carried out in
great detail on chaotic cavities\cite{Oberholzer1} and chains of
cavities\cite{Oberholzer2}.

The measured Fano factors in all cited experiments can be entirely explained
by theories based on the semiclassical Boltzmann-Langevin
formalism\cite{Nagaev3,Rudin1,Beenakker2,Sukhorukov1}, since both Coulomb
blockade and quantum interference were unimportant.  Although the third
cumulant of FCS of a diffusive wire in the hot electron regime was obtained
quantum mechanically\cite{Gefen1}, it is clearly desirable to have a fully
semiclassical method to treat the FCS of semiclassical systems. An important
step in this direction has been undertaken in Ref.~\onlinecite{Nagaev2} where a
diagrammatic scheme for higher order cumulants has been proposed. A theory for
the full charge distribution of FCS based on a stochastic path integral was
then presented in Ref.~\onlinecite{Pilgram1} and extended to time-dependent
correlation functions in Ref.~\onlinecite{Pilgram3}.  It is the aim of this article
to apply the stochastic path integral approach to systems in the hot electron
regime.

This article is organized as follows: The first geometry under consideration,
the chaotic cavity, is introduced in Sec.~\ref{Chaotic Cavity}.  The principal
tools to obtain the FCS for this system are discussed in
Secs.~\ref{Preliminaries} and \ref{Calculations}, results are presented in
Secs.~\ref{Results} and \ref{Frequency Dispersion}. In Sec.~\ref{Diffusive
Wire}, we compare the statistics of cavity and diffusive wire and give
in the conclusions~\ref{Conclusions} some estimates which show that our
calculations are relevant for experiments.

\section{Chaotic Cavity}
\label{Chaotic Cavity}

\begin{figure}[t]
\begin{center}
\leavevmode
\psfig{file=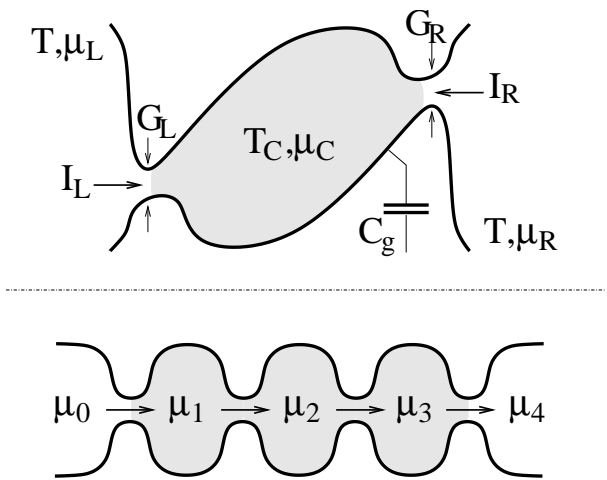,width=8cm}

\caption{The two geometries under consideration in this article. {\it Upper
    panel:} A chaotic cavity in the hot electron regime which is characterized
    by a fluctuating effective chemical potential $\mu_C$ and an effective
    temperature $T_C$. {\it Lower panel:} A chain of $N$ cavities which mimics
     in the large $N$ limit a diffusive wire.}
\label{Sketch Geometries}
\end{center}
\end{figure}

\subsection{the model}
\label{The Model}

The system we study in this section is shown in the upper panel of
Fig.~\ref{Sketch Geometries}.
A chaotic cavity is a conducting island of irregular shape. Its properties are
essentially defined by two quantities, the mean spacing of electron levels
$\Delta = N_F^{-1}$ and the capacitance $C_g$ which describes the Coulomb
interaction of the cavity and its environment represented by a nearby
gate. The island is connected to external leads by two point contacts. These
openings are chosen to be small compared to the size of the cavity in order to
keep motion inside the cavity chaotic. The leads are assumed to be in local
thermal equilibrium described by Fermi distribution functions
$f_{L,R}(\varepsilon)=\{1+\exp[(\varepsilon-\mu_{L,R})/T]\}^{-1}$ (here and in
the following, we set $\hbar=1$, $e=1$ and $k_B=1$).

Depending on the conductances $G_{L,R}$ of the two point contacts, this system
shows despite of its conceptual simplicity a rich variety of physical regimes:
For low dimensionless conductance $G_{L,R}\ll 1$, the charge transport through
the cavity is dominated by Coulomb blockade effects and may exhibit Kondo
physics\cite{Aleiner1}.  
For intermediate conductance $1 \lesssim G_{L,R}$,
weak Coulomb blockade is still possible if temperature and applied voltage
$\mu = \mu_L-\mu_R$ are small enough.
Furthermore, the conductance of the cavity is subject to weak
localization corrections\cite{Beenakker3}. In the semiclassical regime
$G_{L,R}\gg 1$, signatures of the quantum nature of the charge carriers
disappear completely in measurements of mean currents. The total conductance
of the cavity is now simply given by Ohm's law, i.e. by the conductance of the
two point contacts in series ($G=G_LG_R/(G_L+G_R)$). However, even in this
semiclassical limit, the quantum statistics of the electrons has a strong
influence on transport fluctuations, namely on the shot noise produced by the
two point contacts.  For simplicity, we will assume that both contacts are
completely open, i.e. no backreflection is taking place at the point contacts.

In the semiclassical regime, the state of the cavity is characterized by its
time-dependent electron occupation function $f_C(\varepsilon-U_C)$ and the
electrostatic potential $U_C$.  The chaotic scattering inside the cavity
renders this occupation function position independent and
isotropic\cite{Comment Semiclassical}.  Due to the random nature of the
currents flowing into the cavity, the occupation function
$f_C(\varepsilon-U_C)$ is fluctuating around its mean $f_C^0(\varepsilon)$.
There are four time scales which are important for the dynamics of these
fluctuations: The electron-electron scattering time $\tau_{e-e}$, the
inelastic scattering time $\tau_{e-ph}$, the dwell time $\tau_d = R_qN_F$
which is the time an electron spends inside the cavity, and the rc-time
$\tau_{rc}= R_qC_{\mu}$ that describes the relaxation of charged fluctuations
inside the cavity. Here, $R_q=(G_L + G_R)^{-1}$ denotes the charge relaxation
resistance and $C_{\mu}^{-1} = C_g^{-1} + N_F^{-1}$ is the electrochemical
capacitance of the cavity\cite{Buettiker1}.  The mean occupation function
$f_C^0(\varepsilon)$ depends on the relations between the different time
scales: In the {\it cold electron regime }
($\tau_{e-ph}\gg\tau_{e-e}\gg\tau_d$), the energy of every electron passing
the cavity is conserved. Particle current conservation at each energy implies
$f_C^0 = R_q(G_Lf_L + G_Rf_R)$.  Shot noise in this regime has been first
characterized by random matrix theory\cite{Jalabert1} using the scattering
theory of noise\cite{Khlus1,Lesovik1,Buettiker2} and later by a semiclassical
approach\cite{VanLangen1,Blanter2}. Higher cumulants have been calculated
quantum mechanically for open point contacts\cite{Schomerus1} and
semiclassically for arbitrary contacts\cite{Nagaev1}.  In the {\it dissipative
regime} ($\tau_d\gg\tau_{e-ph}$), electrons entering the cavity are in thermal
equilibrium with the surrounding phonon bath and only the energy integrated
particle current through the cavity is conserved.  Noise for this regime has
been calculated from a voltage probe model\cite{VanLangen1} and recently for
low bias in the framework of circuit theory\cite{Bagrets2}. In the {\it hot
electron regime} ($\tau_{e-ph}\gg\tau_d\gg\tau_{e-e}$), both particle current
and total energy current through the cavity are conserved (for theoretical
predictions of noise see Ref.~\onlinecite{Beenakker2,Oberholzer1}).  The
distribution of the electrons in the cavity is described by a Fermi function
$f_C(\varepsilon) = \{1+\exp[(\varepsilon-\mu_{C})/T_C]\}^{-1}$ where both
electrochemical potential $\mu_C$ and local electron temperature $T_C$ are
fluctuating around their mean values $\mu_C^0$ and $T_C^0$.  Notice that there
is no experimental distinction between these three regimes on the level of
mean current. Only noise measurements reveal the type of interactions present
in the cavity.  From now on, we will focus on the hot electron regime.

\subsection{preliminaries}
\label{Preliminaries}

Before we start do describe the calculation of the FCS, we give a brief
summary of definitions we use throughout the paper.  The complete information
about the statistics of current flow $I(t)$ through a cross section of a
conductor is contained in its probability functional $P_I[I(t)]$. More
general, the probability functional to find a certain realization of a
stochastic variable $A(t)$ defined on the interval $[0,\tau]$ can be written
as Fourier transform
\begin{equation}
\label{Stochastic Variable}
P_A[A] = \int {\cal D}\chi_A 
\exp\left\{-i\int_0^\tau \chi_AA + S_A[i\chi_A]\right\}
\end{equation} 
where we introduced the characteristic functional $S_A[\chi_A(t)]$, the
conjugated field $\chi_A(t)$, and a functional integration over the measure
${\cal D}\chi_A$.  An analytic continuation of $S_A[\chi_A(t)]$ to
an imaginary field $i\chi_A$ is used for the Fourier transformation in
Eq.~(\ref{Stochastic Variable}).  Functional derivatives of the characteristic
functional yield the irreducible part of any correlation function
\begin{equation}
\label{Correlation Function}
\langle A(t_1)\dots A(t_n) \rangle = \left.
\frac{\delta^n S_A[\chi_A(t)]}{\delta\chi_A(t_1)\dots\delta\chi_A(t_n)}
\right|_{\chi_A=0}
\end{equation}
in time representation.  Often, one is interested in the behavior of
stationary systems. It is then useful to introduce the spectral function
$C_A^n(\omega_1,\dots,\omega_{n-1})$ which is linked to the Fourier transform
of the correlation function~(\ref{Correlation Function}) by
\begin{equation}
\label{Spectral Function}
\langle A(\omega_1)\dots A(\omega_n) \rangle =
2\pi \delta(\omega_1+\dots+\omega_n) C_A^n(\omega_1,\dots,\omega_{n-1}).
\end{equation}
The full counting statistics of charge transfer as defined by Levitov et
al.\cite{Levitov1} does not keep the entire information contained in the
functional $S_I[\chi_I]$ for current fluctuations, but retains only the
probability distribution of the time integrated current
\begin{equation}
\label{Counting Statistics}
P(Q) = \int d\chi e^{-i\chi Q + S(i\chi)}, \quad
Q = \int_0^\tau dt I(t).
\end{equation}
The characteristic function $S(\chi)$ of FCS is obtained from the complete
functional $S_I[\chi_I]$ by choosing $\chi_I(t) = \chi$. The counting field
$\chi$ is constant in time.  The distribution $P(Q)$ may be connected to the
spectral function of the current $C_I^n$. In the long-time limit for instance,
the cumulants of the FCS $P(Q)$ turn out to be the spectral functions $\tau
C_I^n$ taken at zero frequency.


\subsection{calculations}
\label{Calculations}

In this section, we discuss the essential steps that are necessary to
calculate the transport statistics in the presence of strong electron-electron
scattering. We employ a stochastic path integral formalism which has been
developed in Ref.~\onlinecite{Pilgram1} to study the FCS of semiclassical mesoscopic
conductors.
This formalism is based on the observation that the correlation time of bare
current fluctuations in the point contacts is the shortest time scale in the
problem. This allows us to proceed in two steps: First, we consider the point
contacts as sources of white noise which obey statistics that depend on
external parameters such as the occupation functions of the leads and the
cavity. Second, we identify a set of conserved currents which allow us to
determine the adiabatic time evolution of the external parameters.
To characterize the bare noise of the point contacts we introduce
characteristic functionals $S_j[\chi_j,\xi_j]$ for the current fluctuations in each contact
(see Eq.~(\ref{Stochastic Variable}))
\begin{equation}
\label{White Noise}
S_j\left[\chi_j,\xi_j\right] = \int_0^\tau 
H_j\left(\chi_j,\xi_j,\mu_j,T_j,\mu_C,T_C\right)d\tau
\quad j = L,R.
\end{equation}
The fields $\chi_j$ and $\xi_j$ are conjugated to particle current $I^p_j(t)$
and total energy current $I^e_j(t)$ respectively. The Fourier transform of
$\exp\{S_j\}$ is the probability functional $P_i[I^p_j,I^e_j]$ to find a
certain realization of currents. The white color of the noise is apparent from
the form of Eq.~(\ref{White Noise}) which contains one single time
integration; probabilities at different times are hence independent. The
precise form of $H_j$ must be taken from a quantum mechanical
calculation. With the help of Ref.~\onlinecite{Levitov1}, one finds for open point
contacts\cite{Pilgram1,Pilgram2}
\begin{equation}
\begin{array}{cc}
H_j(\chi_j,\xi_j) = &
 G_j\int d\varepsilon \ln\left[1+f_j(\varepsilon)
\left(e^{\chi_j+\varepsilon\xi_j}-1\right)\right] +\\
\\
&
 G_j\int d\varepsilon \ln\left[1+f_C(\varepsilon)
\left(e^{-\chi_j-\varepsilon\xi_j}-1\right)\right].
\end{array}
\label{Point Contact Generators}
\end{equation}
Since all occupation functions in Eq.~(\ref{Point Contact Generators}) are
Fermi functions, the energy integrals over $\varepsilon$ remain elementary.
The generating functions of the point contacts as a function of temperature
$T_C$ and chemical potential $\mu_C$ of the cavity are given by
\begin{eqnarray}
&&   H_j(\lambda,\xi,V_j,V_C,T_j,T_C) =  \nonumber\\
& &   \frac{\pi^2G_j}{6} 
\frac{2V_j\lambda+T_j\lambda^2+
\left(\left(T_j\right)^2+\left(V_j\right)^2\right)\xi}
{1-T_j\xi}   \nonumber\\
& &  -\frac{\pi^2G_j}{6} 
\frac{2V_C\lambda-T_C\lambda^2+
\left(\left(T_C\right)^2+\left(V_C\right)^2\right)\xi}
{1+T_C\xi}.
\label{Final Contact Generators}
\end{eqnarray}
To shorten the expressions, we have rescaled the chemical potentials $\mu_j =
\pi V_j / \sqrt{3}$ and the fields $\chi_j = \pi \lambda_j / \sqrt{3}$.

Now that we have solved the problem of charge and energy transfer statistics
through each individual point contact, we can proceed to determine the more
difficult statistics for two point contacts in series. To this end, we need to
find conservation laws that set the slow dynamics of the two variables $\mu_C$
and $T_C$ which define the state of the cavity in the hot electron regime. We
therefore combine the conservation of charge $Q_C$ and total energy $E_C$
inside the cavity
\begin{equation}
\label{Conservation Laws}
\dot{Q}_C = I^p_L + I^p_R \qquad \dot{E}_C = I^e_L + I^e_R  
\end{equation}
with expressions that link the conserved quantities to chemical potential and
temperature
\begin{equation}
\label{Link Cavity Parameters}
Q_C = C_{\mu}\mu_C \qquad E_C = 
\frac{\pi^2}{6} N_F\left(T_C\right)^2 + \frac{1}{2}C_{\mu}\left(\mu_C\right)^2.
\end{equation}
These relations include charge screening inside the cavity on the level of the
Thomas-Fermi approximation\cite{Buettiker1}.
The conservation laws~(\ref{Conservation Laws}) are conveniently expressed by
Lagrange multipliers $\chi_C$ and $\xi_C$ which are introduced through
delta-functionals in Fourier representation.  We use
\begin{equation}
\label{Lagrange Multipliers}
\begin{array}{c}
\delta\left[I^p_L + I^p_R - \dot{Q}_C\right]
=\\
\\
\int {\cal D}\chi_C \exp\left\{
-i\int_0^\tau dt \chi_C\left(I^p_L + I^p_R - \dot{Q}_C\right)
\right\}
\end{array}
\end{equation}
for the charge conservation and a similar expression for the energy
conservation. Without incorporating the conservation laws~(\ref{Conservation
Laws}), the probability to find a certain realization of particle and energy
currents is given by the product $P_L[I^p_L,I^e_L]P_R[I^p_R,I^e_R]$, i.e. the
left and the right point contact are independent. We combine this product with
the delta-functional~(\ref{Lagrange Multipliers}) and construct a conditional
probability
\begin{equation}
\label{Conditional Probability}
\begin{array}{c}
P[I^p_L,I^p_R,I^e_L,I^e_R] = \int {\cal D}Q_C  {\cal D}E_C
\delta\left[I^p_L + I^p_R - \dot{Q}_C\right]\\
\\
\delta\left[I^e_L + I^e_R - \dot{E}_C\right]
P_L[I^p_L,I^e_L]P_R[I^p_R,I^e_R]
\end{array}
\end{equation}
which satisfies the current conservation laws. Introducing the characteristic
functionals~(\ref{White Noise}) of the point contacts, we find that the total
characteristic functional for the cavity can be written as stochastic path
integral\cite{Pilgram1}  
\begin{equation}
\label{Stochastic Path Integral}
\begin{array}{c}
e^{\tilde{S}_I[i\chi_I]} = \int {\cal D}Q_C  {\cal D}E_C
{\cal D}\chi_C  {\cal D}\xi_C
e^{S_I[i\chi_I,i\chi_C,i\xi_C,Q_C,E_C]}
\end{array}
\end{equation}
over the action
\begin{equation}
\label{Effective Stochastic Action}
\begin{array}{c}
S_I[\chi_I,\chi_C,\xi_C,Q_C,E_C] = 
\int_0^\tau dt
\bigl(-(\chi_C\dot{Q}_C + \xi_C\dot{E}_C)\\
\\
+ H_L(\chi_C+\chi_I,\xi_C)
+ H_R(\chi_C,\xi_C)
\bigr).
\end{array}
\end{equation}
In the semiclassical regime, this path integral may be evaluated in saddle
point approximation ($\tilde{S}_I = S_I$ at the saddle point). Due to the
adiabatic evolution of the chemical potential $\mu_C$ and the effective
temperature $T_C$ of the cavity, Gaussian corrections are
small\cite{Pilgram1}.  The four saddle point equations get the form
\begin{eqnarray}
\label{Saddle Point Equations}
\dot{Q}_C = \frac{\partial H_L}{\partial\chi_C} + 
\frac{\partial H_R}{\partial\chi_C},\qquad
\dot{E}_C = \frac{\partial H_L}{\partial\xi_C}  + 
\frac{\partial H_R}{\partial\xi_C},\nonumber
\end{eqnarray}
\begin{eqnarray}
\dot{\chi}_C = - \frac{\partial H_L}{\partial Q_C} - 
\frac{\partial H_R}{\partial Q_C},\qquad
\dot{\xi}_C = - \frac{\partial H_L}{\partial E_C} - 
\frac{\partial H_R}{\partial E_C},
\end{eqnarray}
and resemble formally the canonical equations of motion for position and
momentum in mechanics.
The first two equations can be interpreted as continuity equations that
express the change of charge and energy in the cavity in terms of incoming
currents.  The right sides of third and forth equation can be understood as
forces which prevent the saddle point solutions from exploring unlikely
regions of the configuration space.

\subsection{results}
\label{Results}

In this section, we evaluate Eq.~(\ref{Effective Stochastic Action}) in the
zero frequency limit. In this case, we may set all time derivatives in the
action~(\ref{Effective Stochastic Action}) and the saddle point
equations~(\ref{Saddle Point Equations}) to zero.  The external counting field
$\chi_I(t) = \chi$ becomes time independent and is conjugated to the charge
transmitted through the cavity during time $\tau$; i.e. $S(\chi)=S_I[\chi]$
generates the FCS (see Eq.~(\ref{Counting Statistics})).  We are left with a
system of four nonlinear equations~(\ref{Saddle Point Equations}) to be
solved. It turns out that analytical solutions to this system exist. After
inserting these solutions into Eq.~(\ref{Effective Stochastic Action}) we
arrive at the following result for the generating function of FCS which is
valid in the long-time limit ($G\tau\max\{\mu,T\}\gg 1$)
\begin{equation}
\label{Full Generating Function}
\begin{array}{c}
S(\tilde{\chi}) = \frac{\pi^2}{3}G\tau
\Bigl[V\tilde{\chi} + 
\left( \sqrt{\eta^{-2} + \tilde{\chi}^2} -\eta^{-2} \right)\times \\
\\
\left(
T \sqrt{\eta^{-2} + \tilde{\chi}^2 +}
\sqrt{T^2(\eta^{-2}+\tilde{\chi}^2)+V^2+2TV\tilde{\chi}}
\right)
\Bigr].
\end{array}
\end{equation}
The ratio $\eta = \sqrt{G_LG_R}/(G_L+G_R)$ describes the asymmetry of the
cavity and becomes at most $1/2$ in the case of a symmetric cavity;
$G=G_LG_R/(G_L+G_R)$ is the conductance of the cavity. We remind
the reader that we have rescaled the bias $V = \pi\mu/\sqrt{3}$ and the
counting field $\tilde{\chi} = \pi\chi/\sqrt{3}$.
\begin{figure}[t]
\begin{center}
\leavevmode
\psfig{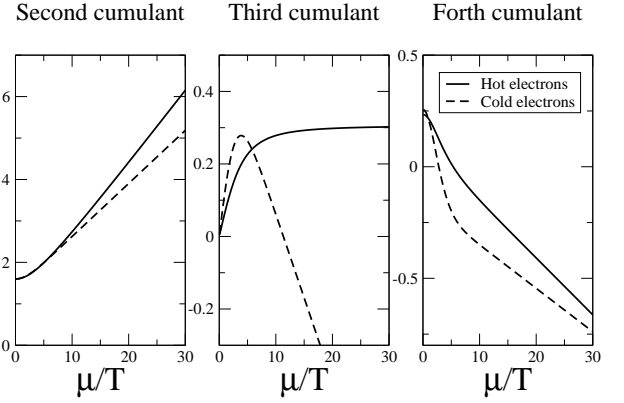}

\caption{Comparison of hot and cold electron regime: The cumulants of counting
  statistics are plotted as a function of the applied bias $\mu$ for an
  asymmetric cavity ($\eta=2/5$). The biggest difference appears in the
  third cumulant which does not change sign in the presence of
  electron-electron scattering. }
\label{Cumulant Comparison}
\end{center}
\end{figure}
Cumulants of transmitted charge $Q=\int_0^\tau dtI(t)$ can be
obtained from derivatives of Eq.~(\ref{Full Generating Function}) with respect
to $\chi$. For the first few cumulants $\tau C_I^n$ one finds
\begin{eqnarray}
C_I^1 = G\mu, \qquad
C_I^2 = G\left(T + T^0_C\right),\nonumber
\end{eqnarray}
\begin{eqnarray}
C_I^3 = \frac{9}{\pi^2}\frac{G^2}{G_L+G_R}
\frac{T\mu}{T^0_C},\nonumber
\end{eqnarray}
\begin{eqnarray}
C_I^4 = \frac{9}{\pi^2}\frac{G^2}{G_L+G_R}
\left(T-T^0_C+\frac{2T^4}{(T^0_C)^3}\right).
\label{First Few Cumulants}
\end{eqnarray}
The mean effective temperature of the cavity is $(T^0_C)^2 = T^2 +
\pi^2\eta^2\mu^2/3$.  The average current $C_I^1$ is obtained from Ohm's law,
the noise $C_I^2$ has been first calculated in
Refs.~\onlinecite{Beenakker2,Oberholzer1} and third and fourth cumulant were
obtained in a perturbative manner in Ref.~\onlinecite{Pilgram1}.  The cumulants are
shown as a function of applied voltage in Fig.~\ref{Cumulant Comparison} and
are compared to the same cumulants in the cold electron regime\cite{Comment
Cold Regime}. The asymmetry is chosen to be $\eta=2/5$.  Several conclusions
can be drawn from Fig.~\ref{Cumulant Comparison} and Eqs.~(\ref{Full
Generating Function},\ref{First Few Cumulants}):
\begin{itemize}
\item
Whereas even cumulants behave qualitatively similar in both regimes, odd
cumulants are strongly suppressed in the shot noise limit of the hot electron
regime. This can be immediately seen in Eq.~(\ref{Full Generating
Function}). There are only two terms which are odd in the counting field
$\tilde{\chi}$. The first term is responsible for the mean current only, and
the second term in the square root is irrelevant for high voltages (shot
noise) and for low voltages (thermal noise).  An asymmetry in the distribution
of transmitted charge thus only appears in an intermediate
regime. Qualitatively, this symmetrization can be explained by the argument
that the hot electron regime is closer to thermal equilibrium where odd
cumulants vanish.
\item
In both regimes, the third cumulant carries the same sign as the first
cumulant for low voltage.  However, for high voltages the third cumulant in
the cold electron regime changes sign.  (Such a sign change does not exist for
two tunnel junctions in series). For cold electrons, the sign change can be
understood from the distribution of transmission eigenvalues. In the hot
electron regime, a careful analysis of cascade
corrections\cite{Nagaev2,Nagaev1} shows that positive correlations between
electron temperature $T_C$ and charge currents are at the origin of this sign.
\item
The transition from low to high bias is smoother in the hot electron regime:
The formulas contain square root laws in contrast to the cold electron regime
where the transition is described by exponential laws.
\item
Even and odd cumulants close to equilibrium are linked by a generalized
fluctuation-dissipation theorem\cite{Sukhorukov2}
\begin{equation}
\label{Fluctuation Dissipation}
2T\frac{\partial C_I^{n-1}}{\partial \mu} = C_I^n \text{  at   } \mu=0.
\end{equation}
\end{itemize}

In the long-time limit, the probability distribution of transmitted charge
which is the Fourier transform of Eq.~(\ref{Full Generating Function}) can be
calculated in saddle point approximation. The upper left panel of
Fig.~\ref{Wire Distribution} shows the probability distribution for low bias,
high bias and for an intermediate regime where the distribution becomes
asymmetric.  An analytical result is available in the shot noise limit. We
find
\begin{equation}
\label{Hot Shot Distribution}
P(Q) \sim \exp\left\{ \frac{\pi\bar{Q}}{\sqrt{3}\eta}
\left(1 - \sqrt{\frac{Q}{\bar{Q}}\left(2-\frac{Q}{\bar{Q}}\right)}\right) \right\},
\end{equation}
i.e. the logarithm of the distribution in the long time limit corresponds to a
semicircle drawn around the mean charge $\bar{Q} = G\tau \mu$.

\subsection{frequency dispersion}
\label{Frequency Dispersion}

In this section, we go beyond the FCS and calculate the frequency dependence
of the third order correlation function~(\ref{Correlation Function}). To this
end, we solve the saddle point equations~(\ref{Saddle Point Equations}) for a
time dependent external field $\chi_I(t)$.  The solution has to be found in a
perturbative manner: First, the response of all internal fields
$\chi_C,\xi_C,\mu_C,T_C$ to the external field $\chi_I$ is calculated up to
quadratic order using the four saddle point equations~(\ref{Saddle Point
Equations}).  Second, the internal fields are inserted into the
action~(\ref{Effective Stochastic Action}).  All terms in the action which are
of third order in the external field will contribute to the desired
correlation function (for details of the calculation see Ref.~\onlinecite{Pilgram3}
where the same calculation for the cold electron regime is discussed
thoroughly).

There is a substantial difference between the noise correlator and higher
order correlation functions which was first pointed out in
Ref.~\onlinecite{Pilgram3}. Dispersion in the noise correlator appears only at high
frequencies given by the inverse rc-time $(\tau_{rc})^{-1}$ and is therefore
difficult to measure. Higher order correlators however, exhibit dispersion
already on a scale given by the inverse dwell time $(\tau_d)^{-1}$ which is a
much lower frequency in metallic systems. Physically, this dispersion is due
to slow fluctuations of the electron temperature $T_C$ which do not show up in
the noise correlator, but couple back into higher cumulants. Such
charge-neutral temperature fluctuations cannot be relieved by particle
currents. The only relaxation mechanism for these fluctuations is heat flow
which reacts on the time scale $\tau_d$.

For the noise correlator in the hot electron regime we find
\begin{equation}
\label{Hot Noise Correlator}
 C_I^2(\omega) = G\left(T + T^0_C\right)
\Delta_2
.
\end{equation}
This is the auto correlation function of currents in the left lead. The
dispersion is described by the expression
\begin{equation}
\label{Noise Dispersion}
\Delta_2(\omega) = 
\Bigl(1+\frac{2\omega^2\tau_{rc}^2}{1-\sqrt{1-4\eta^2}}\Bigr) \Big/
\Bigl(1+\omega^2\tau_{rc}^2\Bigr)
\end{equation}
which depends only on the rc-time $\tau_{rc}$ and the asymmetry parameter
$\eta$.  At low frequencies, $C_I^2$ describes correlated noise of left and
right point contact and reduces to Eq.~(\ref{First Few Cumulants}) for
$\omega=0$. At high frequencies, the statistics of the two point contacts
become independent, $C_I^2$ is then equal to the bare noise of the left
contact.  

The general result for the third order correlation function is too
long to be presented here. We only give the experimentally relevant
behavior at low frequencies ($\omega_{1,2}\ll(\tau_{rc})^{-1}$)
\begin{equation}
\label{Hot Third Correlator}
C_{I,\text{hot}}^3(\omega_1,\omega_2) = 
\frac{9}{\pi^2}\frac{G^2}{G_L+G_R}\frac{T\mu}{T^0_C}
\Delta_3
.
\end{equation}
The dispersion at low frequencies is independent of the asymmetry parameter
$\eta$ and given by
\begin{equation}
\label{Third Dispersion}
\Delta_3(\omega_1,\omega_2) =
\frac{1+\left(\omega_1^2+\omega_2^2+\omega_1\omega_2\right)\tau_d^2/3}
{\left(1 - i\omega_1\tau_d\right)\left(1 - i\omega_2\tau_d\right)
\left(1 + i(\omega_1+\omega_2)\tau_d\right)}.
\end{equation}
It is important to note that charge pile up inside the cavity is forbidden on
time scales longer than the rc-time. Therefore, these current fluctuations are
the same in both point contacts.  

For comparison, we also give the third order correlation function of the cold
electron regime (the zero temperature result has been calculated in
Ref.~\onlinecite{Pilgram3} for more general cavities with backscattering at the
point contacts)
\begin{equation}
\label{Cold Third Correlator}
C_{I,\text{cold}}^3(\omega_1,\omega_2) =
M + \frac{3G^2}{G_L+G_R}\frac{T\sinh(\mu/T) - \mu}{\cosh(\mu/T) - 1}
\Delta_3
\end{equation}
where $M$ is a contribution which is independent of frequency
\begin{eqnarray}
\label{Cold Minimal Correlation}
& &M = \frac{G\left(G_L^2 + G_R^2\right)\left(G_L-G_R\right)^2}
{\left(G_L+G_R\right)^4}\times\nonumber\\
& &\frac{2\mu+\mu\cosh(\mu/T) - 3T\sinh(\mu/T)}{\cosh(\mu/T) - 1}.
\end{eqnarray}
Note that the low-frequency dispersion $\Delta_3(\omega_1,\omega_2)$ of the
third order correlation function~(\ref{Third Dispersion}) is the same as in
the cold electron regime.  Also the prefactors of
$\Delta_3(\omega_1,\omega_2)$ are similar: They both vanish at zero
temperature and at zero bias, i.e. they are proportional to the minimum of
temperature and bias.

\section{Diffusive Wire}
\label{Diffusive Wire}

So far we considered heating effects on current statistics in semiclassical
chaotic cavities. It was particularly easy to introduce the stochastic path
integral formalism to the reader on a geometry which is essentially
zero-dimensional. Moreover, heating effects on noise in this geometry have
been measured\cite{Oberholzer1}. Nevertheless, it is an important question
whether part of our observations do also apply to other systems in which
electron-electron scattering is strong. In fact, noise in the hot electron
regime was first theoretically studied\cite{Nagaev3,Rudin1} and experimentally
verified\cite{Steinbach1,Henny1} in diffusive wires.  It therefore seems
natural to extend our theory to diffusive systems and to compare the results
to the chaotic cavity.

To derive an action similar to Eq.~(\ref{Effective Stochastic Action}) for a
diffusive wire, we consider first a chain of cavities connected by identical
point contacts (see lower panel of Fig.~\ref{Sketch Geometries}).  Every
cavity is capacitively coupled to a close gate.  Such a chain has been studied
experimentally and theoretically in Ref.~\onlinecite{Oberholzer2}.  In a second step,
we take the continuous limit of a large number $N$ of cavities, but keep the
total conductance constant.  The continuous limit corresponds to a diffusive
wire with short range Coulomb interaction (short screening length), since we
have neglected capacitances between neighboring cavities. The action for the
chain is given by
\begin{equation}
\label{Effective Action Chain}
S = \tau\sum_{n=0}^{N-1} H(\lambda_{n+1}-\lambda_n,\xi_{n+1}-\xi_n
,V_n,V_{n+1},T_n,T_{n+1})
\end{equation}
where $H$ of each point contact is defined by Eq.~(\ref{Final Contact
Generators}). Each cavity ($n=1..N-1$) is characterized by its rescaled
chemical potential $V_n=\sqrt{3}\mu_n/\pi$ and its effective temperature
$T_n$. Energy and charge in each cavity obey conservation laws that are
guaranteed by the fields $\lambda_n$ and $\xi_n$. The labels $n=0,N$ belong to
the reservoirs. To calculate the characteristic function $S(\chi)$ of FCS, we
choose the boundary conditions $\xi_0 = \xi_N = \lambda_0 =0$ and $\lambda_N
= \sqrt{3}\chi/\pi$ for the counting fields, $\mu_0=\mu$, $\mu_N=0$ for the
chemical potentials and $T_0=T_N=T$ for the temperature.  The
action~(\ref{Effective Action Chain}) has then to be varied with respect to
all internal fields to obtain $4N-4$ coupled nonlinear saddle point equations.

\begin{figure}[t]
\begin{center}
\leavevmode
\psfig{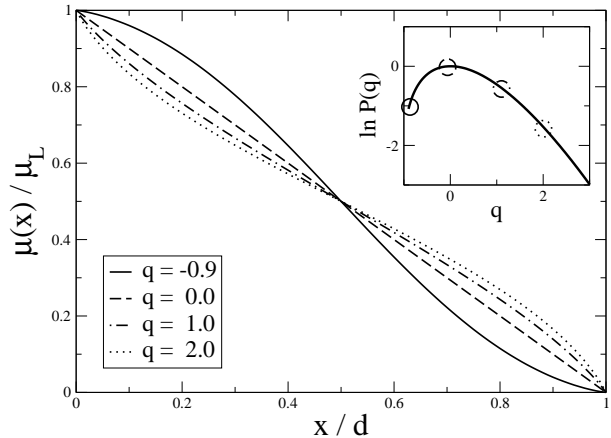}

\caption{Saddle point solutions to the action~(\ref{Field Theory}) of a
  diffusive wire in the hot electron regime: The inset shows the probability
  distribution $P(q)$ of transmitted charge through a wire in the hot electron
  regime in the shot noise limit. The charge is normalized as $q =
  Q/(G\mu\tau)$ and the mean charge has been subtracted. For four selected
  values of $q$, we plot the corresponding saddle point solution for the mean
  chemical potential $\mu(x)$ inside a wire of length $d$.}

\label{Plot Saddle Point}
\end{center}
\end{figure}

In the continuous limit $N\rightarrow\infty$, the difference between
neighboring fields $n$ and $n+1$ becomes infinitesimal and the sum in
Eq.~(\ref{Effective Action Chain}) may be replaced by an integral. We obtain
the action of a nonlinear field theory
\begin{equation}
\label{Field Theory}
\begin{array}{c}
S = \tau \int_0^d dx
\left\{
\left(\begin{array}{cc} \lambda' & \xi'\end{array}\right) \hat{A}
\left(\begin{array}{c} \lambda'\\ \xi'\end{array}\right) +
\left(\begin{array}{cc} \lambda' & \xi'\end{array}\right) \hat{B}
\left(\begin{array}{c} V'\\ T'\end{array}\right)
\right\}.
\end{array}
\end{equation}
The first matrix
\begin{equation}
\label{Field Matrices A}
\hat{A} = 
\frac{\pi^2\sigma}{3}
\left(\begin{array}{cc} T & VT\\ VT & V^2T+T^3\end{array}\right) 
\end{equation}
expresses the noise intensity in local equilibrium. The second matrix
\begin{equation}
\label{Field Matrices B}
\hat{B} = 
\frac{\pi^2\sigma}{3}
\left(\begin{array}{cc} 1 & 0\\ V & T\end{array}\right).
\end{equation}
is the linear response tensor, $\sigma$ denotes the one-di\-men\-sio\-nal
conductance.\cite{General Theory} 

At this point, it seems appropriate to discuss the approximations made in the
derivation of Eq.~(\ref{Field Theory}). Since the action for the diffusive
wire was constructed from the continuous limit of a very specific model, a
chain of cavities linked by open point contacts, one might question the
validity of our ``back of the envelope'' derivation that we employed for
simplicity. It turns out that Eq.~(\ref{Field Theory}) may be obtained as well
from a detailed microscopic calculation based on the Boltzmann-Langevin
approach~\cite{Nagaev3,Rudin1}, if we apply the standard diffusion
approximation and assume the conductor to be quasi-one-dimensional.  Even
without referring to the Boltzmann equation, it can be checked that the
continuous limit of the action~(\ref{Effective Action Chain}) does not depend
on the details of the chain (which may as well contain point contacts of
arbitrary energy independent transparency). In general, the conductance
$\sigma$ then becomes coordinate dependent. We made however an approximation
in neglecting the possible energy dependence of the transparency of the point
contacts. Including such a dependence adds an off-diagonal element
$\hat{B}_{12}\ne 0$ to the linear response tensor, i.e. temperature gradients
then generate a particle flow. Such an additional term implies that the
Wiedemann-Franz law is no longer valid and that the Fano factor of shot noise
becomes non-universal~\cite{Sukhorukov1}. Since there is no trace of such a
non-universality in the experiments on noise in diffusive wires
~\cite{Steinbach1,Henny1} and in the chain of cavities~\cite{Oberholzer2}, we
disregard this additional term.

Note that the action~(\ref{Field Theory}) is quadratic in the fields
$\lambda,\xi$, i.e. in the gradient expansion we lost all informations about
higher order correlators of the elementary noise sources (the point
contacts). This property is not restricted to the hot electron regime, but is
a consequence of the diffusion approximation and the assumption that there are
no long range interactions leading to displacement currents. Furthermore, all
terms in the action~(\ref{Field Theory}) depend only on x-derivatives
$\lambda',\xi'$. Therefore, the saddle point equations which are obtained from
varying Eq.~(\ref{Field Theory}) with respect to $\lambda$ and $\xi$ take the
form of continuity equations for charge and energy conservation. Their
solutions are easily found for the special boundary condition $\chi=0$.
We obtain the profiles of mean chemical potential and mean
effective temperature in the wire
\begin{equation}
\label{Mean Profiles}
\begin{array}{c}
V^0 = V_L\left(1-\frac{x}{d}\right)\\
\\
(T^0)^2 = T^2 + V_L^2\left(1-\frac{x}{d}\right)\frac{x}{d}.\\
\\
\end{array}
\end{equation}
The comparison of this result with Refs.~\onlinecite{Nagaev3,Rudin1} serves as a
check for our calculation.  For general counting field $\chi\neq 0$, the
saddle point equations form a set of coupled nonlinear diffusion
equations. 
We have used the discretized version of the action~(\ref{Effective Action
Chain}) to solve these equations and to obtain the FCS
numerically\cite{Comment Numerics}.

\begin{figure}[t]
\begin{center}
\leavevmode
\psfig{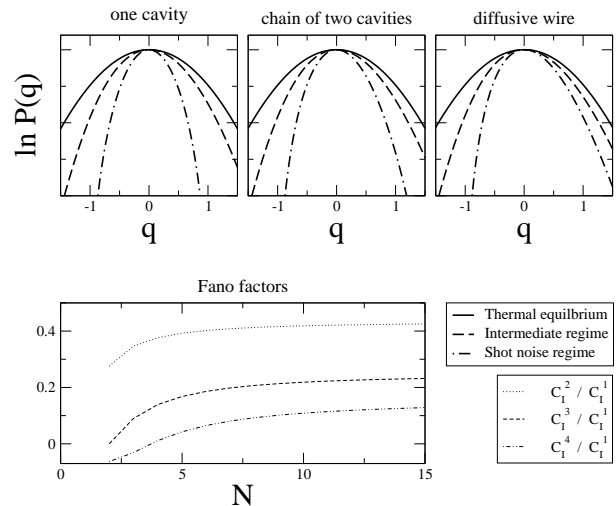}

\caption{{\it Upper panel:} Comparison of counting statistics of chaotic
  cavity, two cavities in series, and diffusive wire in the hot electron
  regime. We use the dimensionless variable $q = Q/(G \max\{\mu,
  \sqrt{3}T/\pi\} \tau)$ to parametrize the charge.  For the intermediate
  case, we use the ratio $\mu/T=2\pi/\sqrt{3}$.  The distribution of charge in
  the shot noise regime is symmetric for a cavity, but asymmetric for a wire.
  {\it Lower panel:} Fano factors of second, third and fourth cumulant in the
  shot noise regime as a function of $N$, the number of point contacts in
  series.}

\label{Wire Distribution}
\end{center}
\end{figure}

Exemplary solutions of the saddle point equations are plotted in
Fig.~\ref{Plot Saddle Point}.  The interpretation of these plots is as
follows: Each shown $\mu(x)$ is the most probable profile of the chemical
potential inside the wire under the condition that the charge $q=Q/(G\mu\tau)$
is measured after time $\tau$. If for instance $q=0$, i.e. the mean charge is
measured, $\mu(x)$ has a linear profile. If the transmitted charge after a
measurement is very small ($q=-0.9$), the potential drop occurs mostly in the
middle of the wire: charges entering from the left therefore most likely
diffuse back to the reservoir they have come from, before they follow the
gradient of the chemical potential.  If the transmitted charge is large
($q=2.0$), the drop occurs mostly close the reservoirs. This means that many
charge carriers from the left side are attracted to the middle of the wire,
and escape to the right.

The upper panel of Fig.~\ref{Wire Distribution} compares the FCS of a cavity,
of a chain of two cavities, and of a diffusive wire in the hot electron
regime. It shows the distributions of transmitted charge $P(q)$ in thermal
equilibrium $\mu\ll T$, for high bias $\mu\gg T$ and for an intermediate case
$\mu / T = 2\pi / \sqrt{3}$.  For zero temperature, the distributions are
bounded from below and except for the wire also from above. For finite
temperatures, the distributions acquire exponential tails.  The most striking
difference between the zero-dimensional cavity and the one-dimensional wire
appears in the shot noise limit in which the distribution for the cavity
becomes symmetric whereas the distribution for the wire stays asymmetric.

The lower panel of Fig.~\ref{Wire Distribution} illustrates the behavior of
the Fano factors at zero temperature $F_n = C_I^n / C_I^0$ of several
cumulants as a function of $N$, the number of point contacts. The result for
$F_2$ is taken from Ref.~\onlinecite{Oberholzer2}.  The results for $F_3$ and $F_4$
are new, in the diffusive limit $N\rightarrow\infty$, $F_3$ agrees with the
literature\cite{Gefen1}.  The increase of $F_3$ towards the diffusive limit
describes the growing asymmetry of the charge distribution, the sign change of
$F_4$ indicates that tails become more important in the diffusive limit.

\section{Conclusions}
\label{Conclusions}

In this article, we presented a method to calculate the full counting
statistics of semiclassical mesoscopic conductors in which electron-electron
scattering is important. This method is based on a separation of time scales:
The correlation time of extraneous sources of noise - such as the point
contacts leading into a chaotic cavity or impurity scatterers in diffusive
wires - is short compared to the characteristic time scales for the evolution
of the electron distribution function. This allowed us to employ a stochastic
path integral for the generating function of counting statistics which can be
solved in saddle point approximation\cite{Pilgram1}.

We derived a variety of results: For a chaotic cavity in the hot electron
regime, the generating function of full counting statistics is given
analytically. For this system we find that the thermalization of the electron
gas in the cavity tends to symmetrize the probability distribution of counted
charge.  Whereas even cumulants are proportional to the maximum of external
temperature and applied bias, odd cumulants are proportional to the minimum
and vanish therefore in the shot noise regime. To the contrary, for a
diffusive wire in the hot electron regime our numerical calculations show that
this symmetrization does not exist.  Furthermore, we discussed the frequency
dependence of the third order correlation function and explained its
unexpected low-frequency dispersion by charge-neutral fluctuations of the
local electron temperature.

We stress the experimental importance of our findings: Noise measurements in
wires with strong electron-electron scattering have been carried out with
great precision\cite{Steinbach1,Henny1}.  The resistances of the samples used
in these experiments range from 1 to 300$\Omega$. The first successful
measurement of a third cumulant\cite{Reulet1} was carried out within this
range.  The third moment of FCS for diffusive wires in the hot electron regime
should therefore be measurable in a similar setup. The chaotic cavities used
in the experiments\cite{Oberholzer1,Oberholzer2} had higher resistances
and would require a different setup to measure the third moment.
The low-frequency dispersion sets in at an inverse dwell time $(\tau_d)^{-1}$
which we estimate from Ref.~\onlinecite{Oberholzer1} to be 1-10GHz.  Dispersion might
therefore be of importance, since the bandwidth used in
experiment\cite{Reulet1} is of the same order.

As a final remark we would like to note that heating effects due to
electron-electron scattering may not only occur inside a cavity or a diffusive
wire, but also inside reservoirs. Most calculations (including the one
presented in this article) assume reservoirs to be at local thermal
equilibrium. This is completely justified for the electrochemical potential,
since the electromagnetic signal propagates quickly throughout the
reservoir. However, fluctuations of the electron temperature can be important
at high frequencies, because their relaxation is much slower. Such temperature
fluctuations in the reservoirs may couple back into higher order correlation
functions of the electrical signal and may cause a low-frequency dispersion
similar to the dispersion that we described in this article.  In such cases,
the notion of an ideal reservoir at local thermal equilibrium has to be given
up.

We would like to thank E. V. Sukhorukov, K. E. Nagaev, and M. Kindermann for
inspiring discussions and acknowledge support from the Swiss National Science
Foundation.


\begin{thebibliography}{2000}
\bibitem{Comment Semiclassical} Here it is implicitly assumed
                                that diffraction is strong, in other words
                                that the Ehrenfest time is much shorter than
                                the dwell time of electrons in the cavity.
\bibitem{Comment Cold Regime} The stochastic path integral approach can be
  applied to the cold electron regime as well. In this case, the occupation
  function $f_C(\varepsilon-U_C)$ fluctuates independently in each energy
  interval $d\varepsilon$ because single particle energy is conserved. For
  more details see Ref.~\onlinecite{Pilgram1}. The curves for Fig.~\ref{Cumulant
  Comparison} were obtained in this way. Alternatively, the cumulants may be
  obtained from a quantum mechanical calculation: The FCS for non-interacting
  electrons in terms of transmission probabilities\cite{Levitov1} may be
  averaged over the distribution of transmission eigenvalues for a open
  chaotic cavity (this distribution was determined in
  Refs.~\onlinecite{Baranger2,Jalabert1} and can be found in
  review~\onlinecite{Beenakker3}).
\bibitem{General Theory} 
   The generalization to higher dimensions is straightforward.
   A similar non-linear field theory can also be derived
   to describe the FCS of a diffusive conductor in the cold electron
   regime.\cite{Jordan1}
\bibitem{Comment Numerics} We use an iterative Newton scheme to solve
                           the nonlinear field equations.
\end{thebibliography}
\end{document}